**High sampling rate single-pixel digital holography system employing a DMD and phase-encoded patterns**

Humberto González, Lluís Martínez-León, Fernando Soldevila, Ma. Araiza-Esquivel, Jesús Lancis, and Enrique Tajahuerce

# High sampling rate single-pixel digital holography system employing a DMD and phase-encoded patterns

HUMBERTO GONZÁLEZ,[1,2] LLUÍS MARTÍNEZ-LEÓN,[1,*] FERNANDO SOLDEVILA,[1] MA. ARAIZA-ESQUIVEL,[2] JESÚS LANCIS,[1] AND ENRIQUE TAJAHUERCE[1]

[1]*GROC·UJI, Institute of New Imaging Technologies (INIT), Universitat Jaume I, Castelló de la Plana 12071, Spain*
[2]*Unidad Académica de Ingeniería Eléctrica, Universidad Autónoma de Zacatecas (UAZ), Zacatecas 98000, Mexico*
*\*lluis.martinez@uji.es*

**Abstract:** A single-pixel digital holography system with phase-encoded illumination using a digital micromirror device (DMD) as a spatial light modulator (SLM) is presented. The enhanced switching rate of DMDs, far exceeding the stringent frame-rate of liquid crystal SLMs, allows recording and reconstruction of complex amplitude distributions in just a few seconds. A single amplitude binary modulation device is used for concurrently displaying the phase-encoded sampling patterns, compensating the distortion of the wavefront, and applying phase-shifting, by means of computer generated holograms. Our detection system consists of a simple photodiode that sequentially records the irradiance fluctuations corresponding to the interference between object and reference beams. The system recovers phase and amplitude information even when a diffuser is placed in front of the photodiode.

## References and links

Research Article                            Vol. 26, No. 16 | 6 Aug 2018 | OPTICS EXPRESS 20343

**Optics EXPRESS**15. N. Savage, "Digital spatial light modulators," Nat. Photonics **3**(3), 170–172 (2009).
16. S. A. Goorden, J. Bertolotti, and A. P. Mosk, "Superpixel-based spatial amplitude and phase modulation using a digital micromirror device," Opt. Express **22**(15), 17999–18009 (2014).
17. D. Stuart, O. Barter, and A. Kuhn, "Fast algorithms for generating binary holograms," arXiv:1409.1841v1 (2014).
18. W. Zhou, Q. Xu, Y. Yu, and A. Asundi, "Phase shifting in-line digital holography on a digital micro-mirror device," Opt. Lasers Eng. **47**(9), 896–901 (2009).
19. R. Horisaki, H. Matsui, and J. Tanida, "Single-pixel compressive diffractive imaging with structured illumination," Appl. Opt. **56**(14), 4085–4089 (2017).
20. P. Clemente, V. Durán, E. Tajahuerce, V. Torres-Company, and J. Lancis, "Single-pixel digital ghost holography," Phys. Rev. A **86**(4), 041803 (2012).
21. S.-H. Zhang, S. Gan, D.-Z. Cao, J. Xiong, X. Zhang, and K. Wang, "Phase-reversal diffraction in incoherent light," Phys. Rev. A **80**(3), 031805 (2009).
22. W. Gong and S. Han, "Phase-retrieval ghost imaging of complex-valued objects," Phys. Rev. A **82**(2), 023828 (2010).
23. T.-C. Poon, "Optical scanning holography: a review of recent progress," J. Opt. Soc. Korea **13**(4), 406–415 (2009).
24. P. Clemente, V. Durán, E. Tajahuerce, P. Andrés, V. Climent, and J. Lancis, "Compressive holography with a single-pixel detector," Opt. Lett. **38**(14), 2524–2527 (2013).
25. L. Martínez-León, P. Clemente, Y. Mori, V. Climent, J. Lancis, and E. Tajahuerce, "Single-pixel digital holography with phase-encoded illumination," Opt. Express **25**(5), 4975–4984 (2017).
26. N. J. A. Sloane and M. Harwit, "Masks for Hadamard transform optics, and weighing designs," Appl. Opt. **15**(1), 107–114 (1976).
27. C. M. Watts, D. Shrekenhamer, J. Montoya, G. Lipworth, J. Hunt, T. Sleasman, S. Krishna, D. R. Smith, and W. J. Padilla, "Terahertz compressive imaging with metamaterial spatial light modulators," Nat. Photonics **8**(8), 605–609 (2014).
28. W. Gong and S. Han, "Correlated imaging in scattering media," Opt. Lett. **36**(3), 394–396 (2011).
29. Z. Zhang, X. Ma, and J. Zhong, "Single-pixel imaging by means of Fourier spectrum acquisition," Nat. Commun. **6**(1), 6225 (2015).
30. W.-H. Lee, "Binary synthetic holograms," Appl. Opt. **13**(7), 1677–1682 (1974).
31. D. B. Conkey, A. M. Caravaca-Aguirre, and R. Piestun, "High-speed scattering medium characterization with application to focusing light through turbid media," Opt. Express **20**(2), 1733–1740 (2012).## 1. Introduction

Different imaging techniques grapple with the challenge of retrieving the complete information from complex samples, including poorly absorbing or weakly scattering objects when no exogenous contrast agents are used. Among them, phase methods allow the qualitative and quantitative evaluation of a myriad of biological specimens [1–4].

Digital holography is a well stablished method for simultaneous phase and amplitude imaging, with straightforward procedures for the reconstruction of complex diffraction patterns [5]. As an interferometric technique, holography provides high sensitivity while meaning a non-invasive procedure. In particular, phase-shifting digital holography is a common approach where a set of interference patterns, with a different phase-shift between object and reference beams have to be recorded [6]. This procedure efficiently exploits the spatial resolution of digital sensors [7].

Single-pixel imaging has focused the interest of a large number of researchers as an innovative computational imaging technique [8]. Instead of using a conventional camera, single-pixel techniques rely on the sequential projection of microstructured light patterns onto a sample while a simple photodiode records the light intensity collected for each pattern. The photocurrent fluctuations corresponding to sampling the scene with these different microstructured masks encompass the spatial information of interest. Thus, the stage of the spatial sampling is shifted away from the sensor, the camera in traditional imaging techniques, to the programmable spatial light modulator where the masks are displayed.

Single-pixel imaging offers a series of advantages in harsh or exotic environments where detectors with a 2D structure are not accessible, like in the spectral bands where spatial high resolution sensors are difficult to build [9], also in applications which are limited by scarce illumination [10] or when vision through scattering media is envisaged [11–13]. Single-pixel techniques are well adapted to apply compressive sensing algorithms, reducing in this way measurement and processing time [14]. Furthermore, the simplicity of the sensor can be exploited to measure new physical imaging dimensions, such as wavelength or polarization, of the sample under scrutiny.

Liquid crystal spatial light modulators (LC-SLMs) are commonly used to codify the sampling patterns in single-pixel techniques, although these phase-modulation devices have a stringent frame-rate limitation, usually around 60 Hz [15]. In contrast, digital micro-mirror devices (DMDs) can reach frame rates up to 22 kHz. A DMD is a reflective type of binary SLM, with multiple tiny switchable mirrors. Each mirror can be switched to either + 12° or −12° with respect to the surface normal, corresponding to the "on" or "off" state, respectively. Even if a DMD is an amplitude modulator, several techniques to modulate phase and amplitude of light beams with this kind of modulators have been reported [16,17]. DMDs have also been used in digital holography systems [18].

Single-pixel imaging techniques with structured illumination have been applied to phase imaging with different approaches such as coherent diffraction imaging [19], ghost imaging [20–22], or digital holography [23–25]. The technique for coherent diffraction imaging is based on compressive sensing [19]. In the case of ghost holography, the object is sampled with random speckle patterns [20] and the phase is reconstructed combining interferometric techniques based on correlations with single-pixel measurements [20–22]. Among the methods based on digital holography, optical scanning holography employs a time-varying Fresnel zone plate, which is displayed onto each point of the object scene, to collect the light diffracted with a photodiode [23]. In the case of single-pixel digital holography, intensity modulated patterns based on Hadamard functions have been employed for recording digital holograms in a single-pixel detection scheme, using compressive sensing techniques for reconstruction [24]. The use of sampling patterns based on functions of a basis, such as the Fourier or Hadamard basis, allows to apply reconstruction techniques different to those employed in ghost imaging, such as direct basis transformations. Recently, single-pixel digital holography using complex-encoded masks, together with phase-shifting techniques, has been reported [25]. In this case, the object is sampled with a set of micro-structured phase patterns, instead of amplitude patterns, implemented onto a liquid-crystal spatial light modulator, while a bucket detector sequentially records interference irradiance fluctuations.

In this paper, we present a high-speed single-pixel phase-shifting digital holography system with phase-encoded illumination. A DMD, integrated in a Mach-Zehnder interferometer, is used for projecting the sampling patterns onto the object. These masks are phase-encoded Hadamard patterns generated by an amplitude modulator codifying computer generated holograms. The detection system consists of a simple photodiode and a digitizing acquisition card recording the integrated irradiance of the interference pattern between the light diffracted by the object and the reference light beam. The high-sampling rate provided by this sort of modulator enables the recording of the complex amplitude distribution of an object at high speed, in a nearly real-time working mode. Moreover, the system is able to measure the aberration induced by the optical system in the absence of a sample object, and to compensate the wavefront in the phase-modulation stage. In our proposal, a single amplitude modulator SLM is used and no additional phase-shifter device is needed. Three different types of phase information are considered: the sampling patterns, the phase shifts to perform on-axis digital holography and the compensation of the wavefront distortion induced by the aberrations of the optical setup. The system is envisaged as an adaptive system, measuring and compensating the wavefront distortions. Our holographic technique benefits from the advantages of single-pixel imaging approaches. Indeed, images of a complex-valued object have been obtained with a 25° holographic diffuser placed more than one centimeter before the photodiode.

The remaining of the paper is structured as follows. Section 2 presents the main features of single-pixel phase imaging. Section 3 introduces the principles of our approach for high sampling-rate complex-valued-encoded single-pixel phase imaging, with an experimental implementation described in Section 4. Experimental results are reported in Section 5, and Section 6 includes the concluding remarks.

## 2. Single-pixel phase imaging

In contrast with conventional imaging techniques using a 2D sensor, single-pixel imaging collects the light intensity codifying the information of interest with a bucket detector, such as a simple photodiode, while the 2D structure is introduced elsewhere in the system. In practice, as different micro-structured patterns are implemented sequentially onto a spatial light modulator (SLM) to sample the scene, photocurrent fluctuations are recorded by a single-pixel sensor. In other words, the purpose of traditional single-pixel imaging is the optical reconstruction of an intensity object, $O(\vec{r})$, through the projection of a set of sampling patterns $p_i(\vec{r})$, with $i = 1,2,…,N$, and the recording of the resulting values of total irradiance. The irradiance transmitted through the sample, $I_i$, for the sampling pattern $i$, is recorded with a bucket detector after the output beam has been focused on it. In mathematical terms, the integrated intensity can be expressed as

$$I_i \propto \iint \left| O(\vec{r}) p_i(\vec{r}) \right|^2 d^2\vec{r}. \tag{1}$$

Afterwards, an intensity object can be retrieved as the linear superposition of the sampling functions weighted with the recorded photocurrents,

$$\left| O(\vec{r}) \right|^2 = \sum_i I_i p_i(\vec{r}). \tag{2}$$

Further, this procedure can be extended to phase measurements with coherent light. When using an interferometer, the output pattern produced by the coherent sum of light coming from the object and from a reference beam may also be integrated by a photodiode. If phase-shifting techniques are applied, several measurements should be performed for each sampling pattern, with a different global phase-shift between the object and the reference beams. The complete reconstruction of the object, in amplitude and phase, will result. For an object $O(\vec{r})$ sampled with patterns $p(\vec{r})$, an amplitude distribution $R(\vec{r})$ in the reference beam, and a phase-shift $\varphi$ between reference and object arms, the irradiance measurements $I_i$ corresponding to each interference pattern can be expressed as

$$I_{i,\varphi} \propto \iint \left| O(\vec{r}) p_i(\vec{r}) + e^{j\varphi} R(\vec{r}) \right|^2 d^2\vec{r}. \tag{3}$$

When applying a four-step phase-shifting technique, four photocurrent values are registered for each sampling pattern, one for each phase-shift ($\varphi = 0$, $\pi/2$, $\pi$ and $3\pi/2$). The measurement process consists in the determination of the coefficients describing the projection of the object scene in the basis of 2D functions defined by the sampling patterns. In our approach, these coefficients associated to each pattern are complex numbers. They can be obtained from the phase-shifted photocurrents according to the following expression, which adapts the conventional phase-shifting algorithm to single-pixel digital holography:

$$y_i = \frac{1}{4|R|} \left[ \left( I_{i,0} - I_{i,\pi} \right) + j \left( I_{i,3\pi/2} - I_{i,\pi/2} \right) \right], \tag{4}$$

where the amplitude of the reference wave is assumed to be a constant. The complex amplitude distribution of the object is retrieved through the linear superposition:

$$O(\vec{r}) = \sum_i y_i p_i(\vec{r}). \tag{5}$$

The object reconstruction can be interpreted as the superposition of the basis functions, weighted by their corresponding complex coefficients. The procedure described so far can be implemented with amplitude or with complex-valued sampling patterns. Previous results of single-pixel digital holography using amplitude Hadamard patterns as sampling functions [24] or, instead, phase Hadamard patterns [25] have been reported. In both cases, the sampling patterns are generated from the Hadamard-Walsh functions, representing modulation in both horizontal and vertical directions, codifying + 1 or 0 values in the amplitude case, and 0 or $\pi$ in the phase mode. Other measuring functions can be employed but Hadamard patterns are easily implemented and provide minimum variance least-squares estimation of the unknown variables [26]. By encoding the sampling patterns in phase, an enhancement of the signal-to-noise ratio would be ideally expected [27], though light efficiency may be restricted by other factors, but also a compact single-pixel phase-imaging system is possible, as the one presented in the following section.

Some properties of our single-pixel phase-imaging technique need to be explained here. In this approach, the reconstructed 2D complex amplitude distribution, as given by Eq. (5), corresponds to the plane sampled by the measuring patterns. If the location where these masks are projected is the sample plane, the object distribution is reconstructed directly, as in conventional holographic imaging. When projecting the sampling phase patterns onto a different plane, the complex amplitude distribution of a Fresnel diffraction pattern would be measured by single-pixel detection. In this case, the retrieved distribution could be propagated with the diffraction formula to find the object distribution, as in conventional Fresnel holography.

Furthermore, single-pixel phase imaging allows the reconstruction of the hologram when a scattering media is located in front of the detector. Image reconstruction is ensured by the high correlation between the spatial sampling patterns displayed over the object and the photocurrent values recorded by a single-pixel detector at the interferometer output. Thus, even if a scattering medium is interposed before the detector, this relation does not change [11,12,28,29]. Our experimental results illustrate this property.

## 3. High sampling-rate complex-valued-encoded single-pixel phase imaging

LC-SLMs have been employed in digital holography for phase-shifting and for active compensation of the wavefront aberrations induced by the optical system, in spite of their stringent frame-rate limitation, around 60 Hz. The sampling rate of LC-SLMs represents a clear restriction for imaging either dynamical or instable scenes. Besides, the effect of phase fluctuations can be completely reduced if the time range of the recording process is decreased from minutes to seconds. DMDs allows a high-sampling rate system, though, in principle, just binary amplitude modulation is possible. However, methods like the proposed by Lee [30] to create a complex hologram codifying amplitude information in a DMD have been proposed, for example for fast and efficient wave-front characterization and shaping [31]. We have adopted a similar approach for creating computer generated holograms, here applied to single-pixel phase imaging with complex-valued sampling masks. In our technique, the phase pattern of interest is combined with a superimposed linear grating. In Eq. (6), we detail the initial amplitude modulation $m$ for implementing Lee's method with the DMD. Function $\varphi$ represents the phase modulation for each pixel considering the addition of the three phase contributions (sampling patterns, aberration correction and constant phase for phase-shifting) and $\beta$ corresponds to the frequency of the superimposed linear grating.

$$m(x,y) \propto [1+\cos((x-y)\beta - \varphi(x,y))]. \qquad (6)$$

The resulting mask is binarized to be displayed in the DMD. A 4-f optical system images the DMD plane onto the sample plane. A filter at the Fourier plane blocks all the light but one of the first diffraction orders, and just the phase information $\varphi$ is transmitted. The separation of the diffraction orders depends on $\beta$. This filtering could restrict the light efficiency of the system, though this is not an issue in our experiment. The pattern encoded in the amplitude hologram is formed as a complex-valued distribution at the object plane. Indeed, the codification may also include amplitude information by locally tuning the duty cycle of the grating, without changing its periodicity. The grating period determines the distance between diffraction orders at the Fourier plane, which should be large enough to avoid superposition of information between different orders, or undesired band-pass filtering.

The complex function included in the hologram and displayed in the DMD may also enable the correction of the wavefront errors present in the illuminating light. Such wavefront errors can be caused by optical aberrations in the lenses, or by the non-planarity of the DMD itself. Indeed, we have determined the aberration induced by the optical system through the measurement in the absence of a sample object. The phase errors produced by the optical system can be compensated further on by projecting sampling patterns including the adequate phase information, or can be corrected in every measurement by realizing an initial calibration of the phase distortion, to be subtracted in the processing stage. Since this process can be performed nearly in real time, with two series of measurements, with and without an object, our optical setup is a true adaptive system.

### 4. Experimental setup

The experimental setup, based on a Mach-Zehnder-type interferometer, is shown in Fig. 1. A fiber-coupled solid state laser (Oxxius-532-50-COL-SLM) with a wavelength of 532 nm was used as light source, and a DMD (DLP Discovery 4200 Texas Instruments), with a pixel pitch of 13.7 μm, as amplitude SLM. After a first beam splitter, the beam illuminating the DMD is filtered at the Fourier plane of a lens with a focal distance of 250 mm. The transmission aperture of the filter has a diameter of 2.4 mm. The one order of diffraction not intercepted propagates through a lens of 120 mm of focal distance. Without the filter, this 4f system would form the image of the DMD plane at the image focal plane of the second lens. Instead, the phase distribution codified in the DMD, the sampling pattern in our case, is displayed at that final plane, where the object of interest is placed. Finally, the sample beam is superimposed to the reference beam, generating an interferogram that is detected by a simple photodiode. The amount of light reaching the detector is filtered for preventing saturation.

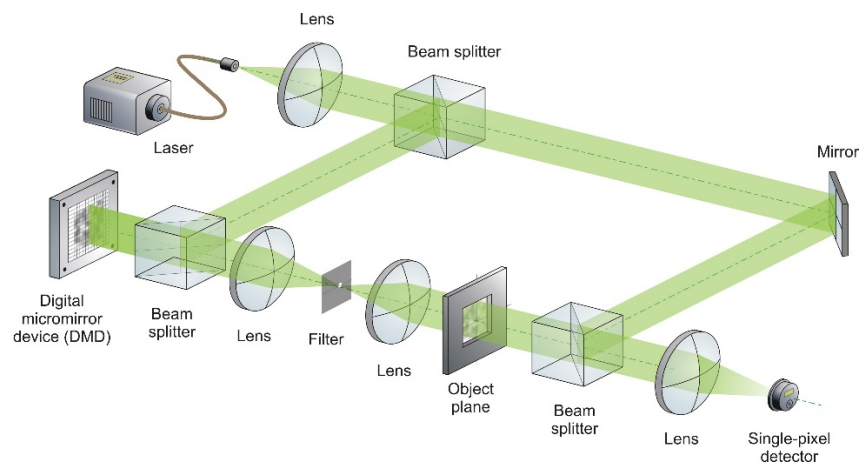

Fig. 1. Experimental setup.

In our measurements, the sampling patterns simultaneously consisted of the Hadamard phase patterns, the relative phase between reference and object beams for phase shifting, and the correction of the optical aberrations previously measured in the system. Hadamard patterns were codified as binary phase masks, with 0 and π values. The resultant total phase modulation was wrapped to 2π. In these experiments, we have worked with an image resolution of 64 × 64 pixels, corresponding to a total amount of 4096 sampling patterns for a standard single-pixel reconstruction. In the DMD, we represented the projection patterns by using 512 × 512 micromirrors. The usual 4-step phase-shifting technique was applied for each one of the patterns displayed, in order to obtain the complex coefficient corresponding to each basis function from the photocurrent value recorded for each phase-shift. A total amount of 16384 images of the computer generated holograms to be displayed in the DMD were created in a preliminary step, and then transferred to the DMD memory.

## 5. Experimental results

Amplitude and phase objects have been measured with the proposed high-sampling rate single-pixel phase-imaging system. The objects have been placed in the plane where the sampling functions were displayed, corresponding to the image plane of the conventional 4f system, where the phase distribution generated with the amplitude modulation of the DMD was obtained. By using this configuration, the phase and amplitude image of the object were directly reconstructed, without any need of numerical propagation. A total number of 4096 images for each one of the four phase-shift values were displayed in the DMD in only two seconds, corresponding to a modulation frame rate of 2 kHz.

As explained above, the beam distortions created in the optical setup can be measured and compensated. The single-pixel phase imaging system is able to calibrate the system-induced aberrations by measuring the wavefront in the absence of an object. Then, it is either possible to project sampling patterns with the correcting phase, as in our experiments, or take these aberrations into account in the reconstruction process. The image of the system-induced aberrations is presented in Fig. 2 (left). Then, the reconstructed image of a complex element, a phase distribution engraved in a photo-resin, when the beam distortions have not been corrected (Fig. 2, center), and when they have been compensated (Fig. 2, right), are shown. A distortion correction of about 3λ is achieved.

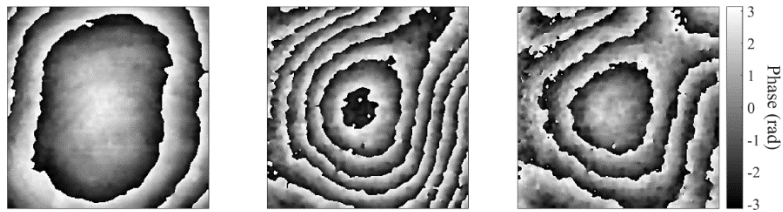

Fig. 2. Phase of the measured system-induced aberrations (left), a distorted measurement of the phase object (center), and the compensated image of the phase object (right).

A transparency with the logo of Universitat Jaume I, has been used as amplitude object, as shown in Fig. 3 (left). Also, a USAF test image registered through our single-pixel digital holography is presented in Fig. 3 (right). The measurements have shown another interesting property of our approach: the object can be retrieved either if the object is placed in the object plane or if it is located in reference beam. In this case, the result is the same provided that the distance from the object to the interferometer output is also identical, as the interference pattern reaching the detector would be similar.

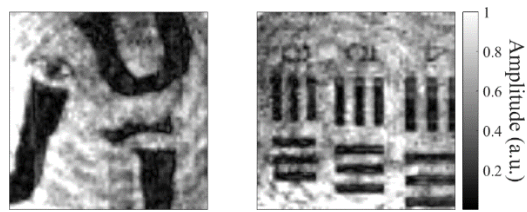

Fig. 3. Reconstruction of amplitude objects: transparency with the Universitat Jaume I logo and USAF test chart.

We have also tested the capability of our system to measure the phase and amplitude of a biological sample. The amplitude and phase of an insect's wing are shown in Figs. 4(a) and 4(c), respectively, when the interferogram is captured by a CCD camera, and Figs. 4(b) and 4(d) show the amplitude and phase of the same object when a photodiode is used as single-pixel detector. For comparison, resolution of images corresponding to digital holography with a CCD has been reduced to 64 × 64 pixels.

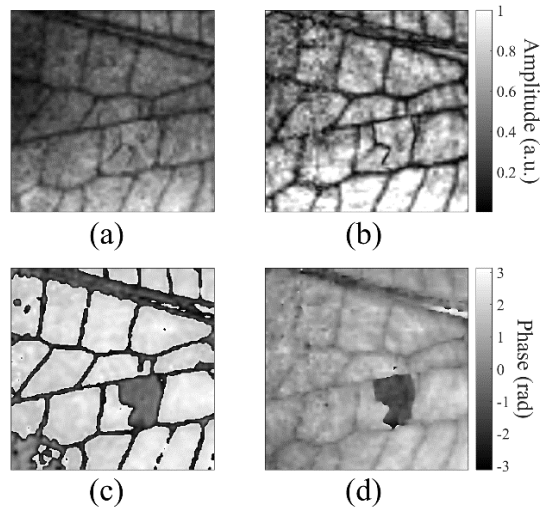

Fig. 4. Phase and amplitude of an insect's wing: (a) amplitude and (c) phase of conventional digital holography images; (b) amplitude and (d) phase of high-speed single-pixel digital holography images.

These images demonstrate the capability of our single-pixel digital holography system to recover complex-valued information employing phase-encoded sampling patterns in just a few seconds. Though the sampling masks with the same Hadamard pattern but different phase-shift were displayed sequentially, a recording time for the whole set of masks of the same phase-shift of less than two seconds can be estimated. Measurement time is not limited by the DMD frame-rate, but by other restrictions related to the software or hardware used in the experiments. Conventional digital holography provides more resolution and, in some cases, a more refined image quality that the performance shown in our proof-of-concept results. But single-pixel digital holography adapts better to applications in which a bucket detector is required or it is convenient for multimodal imaging. Some of the reasons for the limited quality of the single-pixel images are the restricted resolution, the detector performance, or the noise in the digitalization stage, which are just hardware constraints. Signal-to-noise ratio (SNR) calculations have been performed for quantifying quality of amplitude images in Fig. 4: SNR was 12 dB for conventional digital holography and 9 dB for single-pixel digital holography.

Furthermore, we have also tested our system for image reconstruction after transmission of the object information through a scattering medium. A diffuser (Edmund, 25°) has been inserted approximately 1 cm in front of the detector. Figure 5 (left) shows the phase reconstruction of the photo-resin, with a quality comparable to the image without diffuser shown in Fig. 2 (right). In Fig. 5 (right), the phase information of the insect's wing, retrieved when the diffuser is present, is displayed.

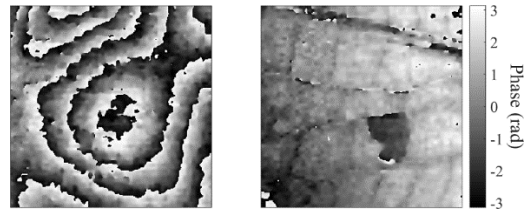

Fig. 5. Phase image of a photo-resin (left) and of an insect's wing (right) when measuring through a scattering medium.

## 6. Conclusions

A single-pixel phase imaging system with complex-encoded illumination at high-sampling rate has been presented. The usual frame-rate limitation of liquid-crystal spatial light modulators is overcome by the use of a DMD. The complex amplitude distribution of an object can be recorded at high speed and reconstructed in just a few seconds. Thanks to the improved robustness of the procedure, high-speed single-pixel digital holography can be achieved. The system is based on phase-shifting techniques and in a method to create phase distributions from amplitude modulation devices. As the micro-structured illumination sampling the object has been codified in phase, a single device can be used for displaying the sampling functions, phase-shifting and compensating the beam distortions introduced by the optical setup. The system acts as an adaptive system, since it is capable of easily measuring and compensating wavefront distortions. Holograms can also be reconstructed when the interferometer output passes through a scattering medium, with a quality comparable to the images recorded directly.

## Funding

Spanish Ministerio de Economía y Competitividad (project FIS2016-75618-R); Generalitat Valenciana (project PROMETEO 2016-079); Universitat Jaume I (project P1·1B2015-35).

## Acknowledgements

One of the authors (H.G.H.) acknowledges partial financial support from CONACYT (Mexico).